\documentstyle[aps,prb,twocolumn]{revtex}

\begin{document}
\author{M. I. Katsnelson$^{1}$ and A. V. Trefilov$^{2}$}
\address{$^{1}$Institute of Metal Physics, Ekaterinburg 620219, Russia\\
$^{2}$Russian Science Center Kurchatov Institute, Moscow 123182, Russia}
\title{Possible scenario of the melting of metals}
\date{\today}
\maketitle
\draft

\begin{abstract}
A microscopic picture of the ``preparation'' of a crystal to the transition
to liquid state at the approach to melting temperature is proposed. Basing
on simple crystallogeometric considerations and the analysis of the
computational results for corresponding anharmonic characteristics an
evaluation is given for the magnitude of atomic displacements leading to the
appearance of close packed Bernal pseudonuclei in the solid phase. A
physical meaning of the classical Lindeman criterion and the mechanisms of
the formation of free volume in the crystal phase are discussed.
\end{abstract}


\narrowtext

Despite a great amount of collected experimental and theoretical information
(see, e.g., \cite{frenkel,skripov,stishov,riemann,brat,zin}) a classical
question what {\it is} the melting is still unsolved. The melting
temperature $T_{m}$ itself is determined from thermodynamic considerations
and, e.g., for alkali metals it can be calculated in the framework of
microscopic theory in a good agreement with experimental data \cite{brat}.
However, it does not elucidate the nature of the processes resulting in the
reconstruction of the structure at the melting point. There are direct
evidences of this reconstruction such as nonlinear in temperature $T$
increase of the heat capacity and a significant softening of shear moduli at 
$T\rightarrow T_{m}$ (for the most of metals the latter is as large as 50\%
decrease of one of the moduli but, e.g., for indium it is even 90\% \cite
{zin}). To investigate the phenomena taking place in the crystal lattices
near $T_{m}$ a broad variety of techniques are used - from classical or
quantum Monte-Carlo computer simulations \cite{gomez,boni} to experiments
with artificial ``crystals'' of charged drops \cite{larsen}. Despite this we
have no yet a complete and reliable picture of the melting of crystals. Here
we try to suppose a qualitative scenario of the melting for a specific case
of metals.

We are interested in the dynamic picture of the processes in a crystal near
the melting point. According to the well-known Lindeman criterion of the
melting one has $\overline{u}/d\simeq 0.1$ at $T=T_{m}$ where $\overline{u}=%
\sqrt{\left\langle u^{2}\right\rangle },$ $\left\langle u^{2}\right\rangle $
being average square of thermal atomic displacements, $d$ is the minimal
interatomic distance (see \cite{brat}). It is important that under such
conditions, as a rule, phonons are well-defined collective excitations in
all the Brillouin zone and anharmonic effects are relatively small up to $%
T_{m}.$ Microscopic calculations for metals with BCC and FCC structures \cite
{anharm} show that the value of relative phonon damping $\Gamma _{{\bf q}\xi
}/\omega _{{\bf q}\xi }$ (where ${\bf q,}\xi $ are the wavevector and phonon
branch number, correspondingly) at $T=T_{m}$ does not exceed 0.1-0.25 for
the most of them. It is naturally to ask why so small anharmonic effects may
lead, nevertheless, to the loss of stability of crystal lattices at the
melting. To our opinion, the answer is connected with geometric properties
of three-dimensional Euclidean space.

Microscopic calculations of structural and thermodynamic characteristics of
liquid rare gases \cite{barker} and liquid metals \cite{brat} show that the
accuracy of their description near $T_{m}$ is dependent mainly on the simple
geometric parameter of atomic packing $\eta =\pi d^{3}n/6$ where $n$ is the
atomic density, $d$ is the diameter of non-overlapped spheres connected with
each atom. It confirms qualitatively the adequacy of geometric approach to
the description of structure of liquids proposed by Bernal \cite{bernal},
namely, the model of random packing of hard spheres. His conclusion about
the existence of close packed ``Bernal pseudonuclei'' formed by the
connection of several tetrahedra \cite{bernal,riemann,skripov} is the most
important for the scenario of the melting proposed here (they are called 
{\it pseudo}nuclei because it is impossible to build regular structure from
these formations). Locally they are characterized by the packing parameter $%
\eta \simeq 0.78$ (at the average packing parameter $\eta \simeq 0.64$)
which is higher than the maximum possible packing in crystals, $\eta \simeq
0.74$ (see \cite{riemann}). To prevent misunderstanding note that we
consider purely geometric definition of the packing parameter corresponding
to the hard-sphere model with the interatomic potential 
\[
\varphi _{HS}\left( r\right) =\{
\begin{array}{l}
\infty ,r<d \\ 
0,r>d
\end{array}
.
\]
Such definition does not take into account the effects of ``softness'' of
interatomic interactions in metals, i.e. the role of the attractive part of
the interactions, as well as nonpairwise forces (the dependence of the
potential on mean electron density) which are important for metals. Because
of this the effective packing parameter in liquid metals at $T\simeq T_{m}$
can be estimated really as $\eta _{met}\simeq 0.45$ \cite{brat}. Despite
this molecular dynamics simulations give us evidences about qualitative
adequacy of the Bernal model of the structure of liquids as a whole \cite
{skripov}.

For metals as well as for liquid rare gases the closest packing is
energetically favorable for small groups of atoms what is confirmed by the
data about the structure of small metallic clusters \cite{marks}. It is well
known that in three-dimensional Euclidean space the closest packing is the
tetrahedral one which, at the same time, cannot be reached in all the space.
This is a drastic difference of three-dimensional space from the
two-dimensional one. The latter can be filled completely by regular
triangles, six of them being meet in each lattice site. In three-dimensional
case only five regular tetrahedra can have the common edge and the void
appears in this case with the angle deficit $\delta =2\pi -5\cos ^{-1}\left(
1/3\right) ,\delta /2\pi \simeq 0.02.$ This simple consideration is the base
of an elegant approach to the description of structure of liquids and
glasses \cite{riemann}. According to this approach the structure is defined
not with the respect to usual Euclidean space but to Riemannian one. The
latter can be, for a specific value of the curvature radius, filled
completely by regular ``tetrahedra''. The transition to the real physical
space is carried out by introducing of corresponding ``decurving'' defects,
namely, structural disclinations. Their appearance near $T_{m}$ has been
recently demonstrated in a simple model of a crystal by molecular dynamics
simulations \cite{jund}. Earlier a similar approach to the problem of
melting basing on the consideration of statistics of linear defects has been
developed phenomenologically by Patashinskii with collaborators \cite{pat}.

Taking into account the elastic energy of the disclinations as well as the
trends to the closest packing of atoms in ``chemical'' energy of interatomic
interactions it may be shown \cite{pack} that the formation of an
inhomogeneous state can be in principle described by these considerations.
This state is characterized by the existence of superdense packed regions
and Bernal ``voids'' around them. The central regions have tetrahedral-like
packing structure and are similar in this sense to Bernal pseudonuclei. Our
main hypothesis is that such regions can be created by thermal fluctuations
near the melting point.

To form the close packed region as a result of thermal motion of an atomic
group it is necessary to ``overcame'' the angle deficit $\delta .$ It has to
be done as a result of not just oscillations but of displacements of the
centers of the oscillations, i.e. of long-lived component of atomic motion.
The latter can be estimated from the following simple considerations. The
irreversibility of atomic displacements results from the phonon damping. The
weight of this irreversible component is of order of $\overline{\Gamma }/%
\overline{\omega }$ where $\overline{\Gamma },\overline{\omega }$ are the
average phonon damping and frequency, correspondingly. The close packed
region will be long-lived enough after its formation by thermal fluctuations
since it corresponds to a local minimum of free energy. Indeed, it is known
that for a system of hard spheres with relatively weak attraction (which is
a good approximation for the most of metals \cite{brat,poten}) the density
of free energy of sufficiently small groups of atoms is minimal for the
closest tetrahedral packing \cite{skripov,marks}. As a result, the
probability of formation of the close packed pseudonucleus can be not too
small for atomic displacements satisfying the condition 
\[
\left( \overline{u}/d\right) \left( \overline{\Gamma }/\overline{\omega }%
\right) \simeq \delta /2\pi \simeq 0.02. 
\]
We believe that this estimation is a realistic melting criterion.
Empirically, it coincides approximately with the Lindeman criterion but, in
contrast with the latter, it is directly connected with contemporary views
about the structure of liquids and a scale of anharmonic effects, at least,
for metals.

Provided that the average atomic density is fixed the formation of close
packed region results inevitably in the appearance of Bernal voids around
it. This changes essentially traditional views formulated by Frenkel \cite
{frenkel} about the leading role of the monovacancies in the formation of
free volume near the melting point. According to the scenario proposed the
latter appears first with the temperature increase as the voids around close
packed regions. Molecular dynamics simulation shows that for systems with
``soft'' enough interatomic interactions (e.g., liquid metals) there are two
kinds of Bernal voids: distorted tetrahedral and octahedral ones \cite
{skripov}. Apparently these voids are the most energetically profitable
``carriers'' of the free volume in the solid phase also. Monovacancies
should have higher free energy and appears only as a result of diffusion
decay of the voids in the close vicinity of $T_{m}.$

To check this assumption we have made the fitting of defect contributions to
the heat capacity of sodium near $T_{m}$ which can be obtained from the
experimental data by substraction electronic, phonon and anharmonic
contributions \cite{VKT}. Trying to fit this contribution $C^{d}\left(
T\right) $ by Arrhenius law in a broad temperature range $\left(
0.6T_{m}<T<T_{m}\right) $ we find that the decline of the curve has a jump
at $T\simeq 0.8T_{m}$. The fitting for the narrower temperature range $%
0.8T_{m}<T<T_{m}$ gives the following results for the entropy $S_{1v}$ and
energy $E_{1v}$ of the formation of monovacancies as well as their
concentration at the melting point $n_{1v}$: $S_{1v}\simeq 4.3,E_{1v}\simeq
0.3$ eV, $n_{1v}\simeq 0.3\%.$ For another temperature range $%
0.6T_{m}<T<0.8T_{m}$ one has $S_{1v}\simeq 0.5,E_{1v}\simeq 0.17$ eV, $%
n_{1v}\simeq 0.74\%$ \cite{VKT}. The first set of the results is in much
better agreement with the data obtained by differential dilatometry data $%
S_{1v}\simeq 5.2,E_{1v}\simeq 0.35$ eV, $n_{1v}\simeq 0.3\%$ \cite{adlhart}
as well as the results of microscopic calculations of the energy of
formation of monovacancy, $E_{1v}\simeq 0.3$ eV \cite{vacancy}.

This confirms that, in contrast with traditional views and in agreement with
the scenario of the melting proposed here, the temperature increase leads
first to the formation of clusters of vacancies (the voids) and the
monovacancies appears only in the close vicinity of $T_{m}.$ Note also that
the formation of monovacancies in a broad temperature range below $T_{m}$
corresponding to Frenkel views about the nature of the melting was never
observed in computer simulations by molecular dynamics \cite{skripov} and
Monte-Carlo \cite{gomez} methods. At the same time, according to these
simulations, pentagonal faces of Voronoi polyhedra are observed frequently
for instant atomic configurations not only in liquids but also in crystals
near the melting point \cite{skripov,gomez}. According to Bernal this is the
main feature of the structure of liquids. It means that some structural
elements which are characteristic for liquids arises by thermal fluctuations
already in the crystal phase.

As a conclusion, consider a possible way to describe such fluctuations
quantitatively. The formation of the closest tetrahedral packing near $T_{m}$
has to lead to the appearance of icosahedral short-range order \cite{riemann}
describing by the parameter 
\[
Q_{6m}\left( {\bf r}\right) =Y_{6m}\left[ \theta \left( {\bf r}\right)
,\varphi \left( {\bf r}\right) \right] 
\]
where $Y_{6m}$ are spherical functions, $\theta \left( {\bf r}\right)
,\varphi \left( {\bf r}\right) $ are polar angles of the direction to the
nearest neighbors of the atom at point ${\bf r}$ and the corresponding
correlation function \cite{icos} 
\[
G\left( r\right) =\frac{4\pi }{13}\sum\limits_{m}\left\langle Q_{6m}\left( 
{\bf r}\right) Q_{6m}^{*}\left( 0\right) \right\rangle . 
\]
Its calculation in the crystal phase by molecular dynamics or Monte-Carlo
methods could give an important information about the local structure of
solids near the melting.

This work is supported by Russian Basic Research Foundation, grant
98-02-16219.

\end{document}